\documentclass{aa}

\usepackage{calc}
\usepackage{amssymb}
\usepackage{amsmath}

\newcounter{subequation}
\renewcommand{\theequation}{\arabic{equation}\ifnum\thesubequation>0{\alph{subequation}}\fi}
\newcommand{\subnumbers}{\setcounter{subequation}{1}}
\newcommand{\nosubnumbers}{\setcounter{subequation}{0}}
\newcommand{\stepsubnumber}{\addtocounter{subequation}{1}\addtocounter{equation}{-1}}
\nosubnumbers

\begin{document}
\thesaurus{02.07.1; 02.08.1; 02.09.1; 03.13.1; 12.03.4; 12.12.1}

\title{Lagrangian theory of structure formation\\
in pressure--supported cosmological fluids}
\author{Susanne Adler \and Thomas Buchert} 

\offprints{Buchert (buchert@stat.physik.uni-muenchen.de)}
\institute{
Theoretische Physik, Ludwig--Maximilians--Universit\"at, 
Theresienstr. 37, D--80333 M\"unchen, Germany}

\date{Received 1 July 1998 / Accepted 6 November 1998}

\titlerunning{Lagrangian theory of structure formation}

\maketitle

\begin{abstract}

The Lagrangian theory of structure formation in cosmological fluids,
restricted to the matter model ``dust'', provides successful models
of large--scale structure in the Universe in the laminar regime,
i.e., where the fluid flow is single--streamed and ``dust''--shells
are smooth. Beyond the epoch of shell--crossing a qualitatively
different behavior is expected, since in general anisotropic
stresses powered by multi--stream forces arise in collisionless
matter. In this paper we provide the basic framework for the
modeling of pressure--supported fluids, restricting attention
to isotropic stresses and to the cases where pressure can be given
as a function of the density. We derive the governing set of
Lagrangian evolution equations and study the resulting system using
Lagrangian perturbation theory. We discuss the first--order
equations and compare them to the Eulerian theory of gravitational
instability, as well as to the case of plane--symmetric collapse. We
obtain a construction rule that allows to derive first--order
solutions of the Lagrangian theory from known first--order solutions
of the Eulerian theory and so extend Zel'dovich's extrapolation idea
into the multi--streamed regime.  These solutions can be used to
generalize current structure formation models in the spirit of the
``adhesion approximation''.

\keywords{Gravitation; Hydrodynamics; Instabilities; Methods: analytical;
Cosmology: theory; large--scale structure of Universe}

\end{abstract}

\vspace{-0.2cm}

\section{Introduction}

Zel'dovich (1970, 1973) initiated the use of Lagrangian coordinates
for the construction of analytical models of cosmic structure
formation. He pointed out that the solution of a force--free continuum
of ``dust'' (i.e. pressureless and non--gravitating matter), given in
terms of trajectories of continuum elements (see, e.g., Zel'dovich \&
Myshkis 1973), can be rescaled to give the correct solution for
gravitational instability in the linear limit. He suggested to follow
those trajectories further up to the epoch when shell--crossing
singularities in the fluid flow develop. Doroshkevich et al.
(1973) confirmed the validity of Zel'dovich's approximation also
within the full self--gravitating system of equations.  To be more
precise they performed a self--consistency check in one of the four
Newtonian field equations (see Buchert 1989 for a discussion and
extension of this self--consistency argument). It was also known
(Zentsova \& Chernin 1980) that Zel'dovich's ansatz provides a
subclass of exact solutions in the case of plane--symmetric collapse
on a Friedmann background.  In (Buchert 1989) it was then shown (going
back to an earlier work on the fully Lagrangian formulation of the
basic equations by Buchert \& G\"otz 1987) that his ansatz also
provides a subclass of exact 3D solutions for a special class of
initial gravitational potentials that are composed of surfaces of
vanishing Gaussian curvature. These solutions feature maximally
anisotropic, locally one--dimensional collapse supporting Zel'dovich's
original discussion of the formation of ``pancakes'' as collections of
volume elements which degenerate into small sheets and later enclose a
three--stream system of the flow. These findings may be restated in
the framework of a Lagrangian perturbation theory in which
Zel'dovich's model is recovered as a subclass of the first--order
solutions (Buchert 1992; for reviews see Bouchet et al. 1995, Buchert
1996, Sahni \& Coles 1996, Ehlers \& Buchert 1997 and references
therein).

The formation of singularities (Arnol'd et al. 1982) is a consequence
and drawback of the matter model which allows fluid elements to
overtake freely. On the other hand, N--body simulations show (e.g.
Doroshkevich et al. 1980) that N--stream systems create additional
multi--stream forces which hinder the central fraction of fluid
elements to escape from high--density regions; as a result they fall
back onto the central region performing oscillations about the center:
an internal hierarchy of nested caustics is formed that continues to
create new structures down to smaller scales (Gurevich \& Zybin 1995).

A phenomenological model based on Burgers' equation was then suggested to
overcome the drawback of ``dust'' models by inventing a Laplacian
forcing that acts in the desired way (Gurbatov et al. 1989).  This
model features ``adhesion'' of already formed structures and so models
the action of self--gravity of multi--stream systems.  A recent
suggestion of how to derive this ``adhesion approximation'' from
kinetic theory of self--gravitating collisionless systems shows that it
is indeed possible to obtain adhesion--type models by taking into
account multi--stream forces (Buchert \& Dom\'\i nguez 1998). However, these
models form a wider class in the sense that multi--stream
forces are, in general, anisotropic unlike in the case of Burgers'
equation. Moreover, equations involving a Laplacian are expected for
the gravitational field strength and not, as is the case for the
``adhesion model'', for the velocity. 

Inspite of the insight gained by deriving the ``adhesion model'' from
first principles, we shall, in the present work, adopt the restricting
assumptions that the multi--stream force acts isotropically (in the
sense of an idealization), and that the dynamical pressure due to
multi--stream systems can be represented as a function of the density.
For these assumptions we derive the general Lagrangian evolution
equations and expand them to first order using the Lagrangian
perturbation approach.
 
A comment concerning the isotropic approximation of the multi--stream
force is in order: our investigation covers pressure--supported fluids
as they are studied in hydrodynamics. Here we do not restrict
ourselves to the description of matter models which are supported by the
usual thermodynamical pressure (i.e., by short--ranged interactions).
Rather we follow an approach that we consider to be relevant for
collisionless systems: a dynamical pressure force arises due to the
action of a ``multi--dust'' region on the bulk motion.  As remarked
above such a forcing is in general anisotropic and the evolution
equations governing the anisotropic stresses involve all velocity
moments of the distribution function (Buchert \& Dom\'\i nguez
1998). Hence, a closure condition is needed which, especially for
self--gravitating fluids (e.g. Bertschinger 1993), is not obvious and
cannot be formulated along the lines of collisional systems. At the
current stage of knowledge about this subtle problem we advance the
phenomenological closure condition that multi--stream stresses may be
idealized by a scalar function \(p\) that is given in terms of the
density of the continuum. This opens the possibility to immediately
access the multi--streamed regime phenomenologically. For example,
Buchert \& Dom\'\i nguez (1998) found that the assumption of
small velocity dispersion singles out a relationship
$p\propto\varrho^{5/3}$ that is solely based on the dynamical
equations; the matter model $p\propto\varrho^2$ yields, together with
the restriction to parallelity of peculiar--velocity and
--acceleration, the ``adhesion approximation'' (Gurbatov et al. 1989).
Other relationships are put into perspective by Buchert et al. (1999);
particular matter models may lead to soliton states which,
due to their persistence in time, could dominate the architecture of
cosmic structure.

\noindent
Although the isotropic idealization is simple--minded, it may actually
be a good approximation in practice, as has been recently found for
high--density regions in numerical simulations (Colombi,
priv.comm.).

\bigskip

We proceed as follows: In Sect.~2 we develop the Lagrangian theory for
pressure--supported fluids and illuminate the structure of the basic
equations by the plane--symmetric case; in Sect.~3 we then move to
some detailed studies of the basic equations: a perturbative treatment
allows to derive a first--order evolution equation for fluid displacements. 
Note that this linear equation in Lagrangian space embodies Eulerian 
nonlinearities. This allows us to formulate an extrapolation rule that extends
solutions of the Eulerian theory of gravitational instabilities into
the nonlinear and multi--streamed regime. This rule rests on the
spirit of Zel'dovich's extrapolation idea that led to the celebrated
``Zel'dovich approximation'' (Zel'dovich 1970, 1973).  We also discuss
the special case where the gravitational collapse is plane--symmetric
on an isotropic background cosmology.  We summarize the results
and provide an outlook in Sect.4.

\vspace{-0.1cm}

\section{The Lagrange--Newton--System with pressure}

\subsection{The Euler--Newton--System with pressure}

Eulerian coordinates are the most widely employed ones in physics. Here we use
\(\vec x\) to define the position that a fluid element occupies at the
moment \(t\) in a non--rotating Eulerian coordinate system. If we
consider \(\vec x\) and \(t\) as independent variables, the total time
derivative of any vector field \( \vec a =\vec a (\vec x ,t)\) equals:
\begin{equation}
\label{totaltimederivate}
\frac{{\mathrm{d}} \vec a }{{\mathrm{d}}t}=\frac{\partial \vec a }{\partial t}+(\vec v \cdot \vec\nabla )\, \vec a \quad .
\end{equation}
The field \( \vec v (\vec x ,t) \) corresponds to the mean velocity in
a kinetic picture. Eulerian coordinates are used to describe fields,
e.g. the spatial mass distribution \(\varrho (\vec x, t)\). In this
description we are not interested in the fluid elements that produce
this density and we do not know their trajectories. In Subsect.2.2 we
introduce Lagrangian coordinates which convey this information.

The basic system of equations describing a self--gravitating medium is
the so--called Euler--Newton--System. We add the accelerative term 
$-\frac{\nabla p}{\varrho}$, which takes into account the
isotropic part of the velocity dispersion tensor. We have the
following evolution equations for the mass density \(\varrho \) and
the velocity \( \vec v \) (\(\vec b \) denotes the acceleration
field): \subnumbers
\begin{gather}
  \frac{\partial \varrho }{\partial t}=-\vec\nabla \cdot (\varrho \vec v ) \label{EulerNewtona}\quad , \\
  \frac{\partial \vec v }{\partial t}=-(\vec v \cdot \vec\nabla )\,
  \vec v +\vec b \quad , \,\hbox {where}\,\, \vec b:=\vec g
  -\frac{\vec\nabla p}{\varrho } \label{EulerNewtonb} \quad .
  \stepsubnumber
\end{gather}
\nosubnumbers The gravitational field strength \(\vec g\) is a
solution of the (Newtonian) field equations:
\addtocounter{equation}{1} \subnumbers
\begin{eqnarray}
  \stepsubnumber \stepsubnumber
  &&\vec\nabla \times \vec g =  \vec 0  \label{EulerNewtonc}\quad ,\\
  \stepsubnumber &&\vec\nabla \cdot \vec g = \Lambda -4\pi G\varrho
  \label{EulerNewtond} \quad ,
\end{eqnarray}
\nosubnumbers where \(\Lambda \) is the cosmological constant and
\(G\) the Newtonian gravitational constant.  The evolution equation
for \(\varrho\) represents the conservation of mass (continuity
equation). The second equation (for \(\vec v\)) is Euler's equation
and describes the balance of momentum. In a kinetic picture
(\ref{EulerNewtonb}) is known as Jeans' equation.

To complete the Euler--Newton--System with pressure we need an
``equation of state'' that relates \(p\) with the dynamical variables
\(\varrho \) and \(\vec v\): we employ the assumption \(p=\alpha
(\varrho) \) with \( \alpha ' :=\frac{\partial\alpha } {\partial
  \varrho }\). Hence, the field equations become
\begin{subequations}
\begin{eqnarray}
\nonumber
\vec 0 =\vec\nabla \times \vec g& =&  \vec\nabla \times \vec b +
\vec\nabla \times \frac{\vec\nabla p}{\varrho } \\
\nonumber &=&\vec \nabla \times \vec b
+ \frac{1}{\varrho}\vec \nabla \times (\vec \nabla p)+\frac{1}{\varrho^2}\vec 
\nabla p\times \vec \nabla \varrho \\
&=& \vec\nabla \times \vec b \quad ; \label{EulerNewtonwithRhoa} \\
\nonumber \Lambda -4\pi G\varrho& =&\vec\nabla \cdot \vec g =  
\vec\nabla \cdot \vec b +\vec\nabla \cdot \frac{\vec\nabla p}{\varrho }\\
\nonumber &=&
\vec\nabla \cdot \vec b +\vec \nabla p \cdot \vec \nabla \frac{1}{\varrho}+
\frac{1}{\varrho}\Delta p\\
&=&  \vec\nabla \cdot \vec b +(\alpha ''-\frac{\alpha '}{\varrho })
\frac{(\vec\nabla \varrho )^{2}}{\varrho }+\frac{\alpha '}{\varrho }
\Delta \varrho \quad \label{EulerNewtonwithRhob} .
\end{eqnarray}
\end{subequations}
The equation for the curl of \(\vec g\) does not change with the
additional accelerative term for that particular class of ``equations
of state''.

\subsection{Transformation tools}

Lagrangian coordinates are assigned to fluid elements and do not
change along the flow lines: we choose the positions \(\vec X\) of the
elements at a time \(t_{0}\) as coordinates (\(\vec X=\vec x\vert_{t_{0}}\)); 
thus we have indexed each fluid element. If we use \(\vec
X\) and \(t\) as independent variables, the total time derivative
becomes \(\frac{\mathrm{d}}{{\mathrm{d}}t}=\frac{\partial}{\partial
  t}\vert _{\vec X}\), i.e., the convective term disappears in this
description.  In order to connect the Lagrangian coordinates with the
Eulerian ones, we introduce the position vector field \(\vec f\),
\begin{equation}
\label{definitionLagrange}
\vec x =\vec f (\vec X ,t)\; ,\quad {\mathrm{where}}\quad \vec X :=\vec f (\vec X ,t_{0})\quad .
\end{equation}

For later convenience we introduce the symbols ``\(\,,\,\)'' and
``\(\,|\, \)'' in order to distinguish Eulerian and Lagrangian
differentiation, e.g. \( \frac{\partial v_{a}}{\partial
  x_{b}}=:v_{a,b} \) and \( \frac{\partial f_{a}}{\partial
  X_{b}}=:f_{a|b} \). \(\vec \nabla \) denotes the Eulerian and \(\vec
\nabla _{\vec 0}\) the Lagrangian nabla operator. In the following we also
apply the summation convention.

The Jacobian matrix, which reflects the deformation of a volume
element, is of key importance for the transformation of coordinates.
Here we use the notation:
\begin{equation}
\label{definitionJ}
J:= \hbox {det}J_{ik}=\hbox {det}f_{i|k}\quad .
\end{equation}
In order to change the Eulerian differentiation to one with respect to
Lagrangian coordinates, we have to define the inverse transformation
of (\ref{definitionLagrange}):
\[ \vec X =\vec h (\vec x ,t)\quad ,\quad \vec h \equiv \vec f ^{-1}\quad.\]
Remembering the definition of the adjoint matrix, \(
{\mathrm{ad}}J_{jk}=\frac{1}{2}\varepsilon _{klm} \varepsilon
_{jpq}f_{p|l}f_{q|m} \), we get
\begin{equation}
h_{k,j}=J_{jk}^{{-1}}=\frac{1}{J}{\mathrm{ad}}J_{jk}=\frac{1}{2J}
        \varepsilon _{klm}\varepsilon _{jpq}f_{p|l}f_{q|m}\quad ;
\end{equation}
\( \varepsilon _{klm} \) denotes the totally antisymmetric Levi-Civita
tensor; note the useful relation \( \varepsilon _{ijk}\varepsilon
_{ilm}=\delta _{jl}\delta _{km}- \delta _{jm} \delta _{kl} \)
(\(\delta _{ij}=1 \) for \( i=j \) and \( \delta _{ij}=0 \) else).  In
order to shorten the equations it is convenient to use functional
determinants:
\begin{equation}
\nonumber
{\cal J}(A,B,C):=\frac{\partial (A,B,C)}{\partial (X_{1},X_{2},X_{3})}:=\left| \begin{array}{ccc}
\frac{\partial A}{\partial X_{1}} & \frac{\partial B}{\partial X_{1}} & \frac{\partial C}{\partial X_{1}}\\
\frac{\partial A}{\partial X_{2}} & \frac{\partial B}{\partial X_{2}} & \frac{\partial C}{\partial X_{2}}\\
\frac{\partial A}{\partial X_{3}} & \frac{\partial B}{\partial X_{3}} & \frac{\partial C}{\partial X_{3}}
\end{array}\right|\end{equation}
\begin{equation}
\label{definitionfunctionaldeterminants}
 =\varepsilon _{klm}A_{|k}B_{|l}C_{|m}\, .
\end{equation}

Because of the properties of the Levi--Civita tensor, or the
definition of a determinant as a multilinear and alternating map,
respectively, we have some tools that make work easier, e.g.:
\begin{eqnarray}
\nonumber
{\cal J}(A+D,B,C)&=&{\cal J}(A,B,C)+{\cal J}(D,B,C) \\ \nonumber
{\cal J}(A,B,C)&=&-{\cal J}(A,C,B) \\ \nonumber
{\cal J}(A,A,C)&=&0 \\ \nonumber
{\cal J}(A\cdot D,B,C)&=& D\cdot {\cal J}(A,B,C)+A\cdot {\cal J}(D,B,C)\,.
\end{eqnarray}
Therefore, we can transform any tensor \( a_{i,j}\) as follows:
\begin{eqnarray}\nonumber
a_{i,j}&=& a_{i|k}h_{k,j}=a_{i|k}\frac{1}{2J} 
\varepsilon _{klm}\varepsilon _{jpq}f_{p|l}f_{q|m} \\
\label{transformation}
&=& \frac{1}{2J}\varepsilon _{jpq}{\cal J} (a_{i},f_{p},f_{q})\quad .
\end{eqnarray}
For further details on Lagrangian evolution equations the reader may
consult the introductory tutorial of Buchert (1996) and the review by
Ehlers \& Buchert (1997).

\subsection{The Lagrange--Newton--System with pressure}

Starting with the Euler--Newton--System (with pressure) we obtain the
Lagrange--Newton--System with the help of the transformation \(\vec
x=\vec f(\vec X,t)\).
 
As the continuity equation (\ref{EulerNewtona}) represents the
conservation of mass \(M=\int _{D_{t}}\varrho {\mathrm d}^{3}x=\int
_{D_{t_{0}}}\varrho J{\mathrm d}^{3}X\) within any domain \(D_{t}\)
that may change in time, i.e.,
\begin{eqnarray}
\frac{{\mathrm d}M}{{\mathrm d}t}=0&=&\frac{{\mathrm d}}{{\mathrm d}t}
\int _{D_{t_{0}}}\varrho J{\mathrm d^{3}X}=
\int _{D_{t_{0}}}\frac{{\mathrm d}}{{\mathrm d}t}(\varrho J){\mathrm d^{3}X}
\nonumber \\
&=&\int _{D_{t}}\frac{\frac{{\mathrm d}}{{\mathrm d}t}(\varrho J)}{J}
{\mathrm d^{3}x}\, ,
\nonumber
\end{eqnarray}
we conclude with $\varrho _{0}:=\varrho(\vec X,t_0)$, $J_{0}:=J(\vec X,t_0)$:
\begin{equation}
\label{varrho0varrhoJ}\varrho J=:C(\vec X)=\varrho _{0}J_{0}\quad. 
\end{equation}
For our definition of Lagrangian coordinates
(\ref{definitionLagrange}) we have \(J_{0}=\det \frac{{\mathrm
    d}x_{i}}{{\mathrm d}X_{j}}\vert_{t_{0}}=1\); nevertheless we will
use \(C(\vec X)\) for the time being in order to allow for a later
relabelling of trajectories.

According to the definition of \( \vec f \) the Eulerian equation
(\ref{EulerNewtonb}) reads:
\begin{equation}
\label{ddotvecf}\vec b =\ddot{\vec f}(\vec X ,t)\quad .
\end{equation}
        
As we have the exact integrals (\ref{varrho0varrhoJ}, \ref{ddotvecf}),
\(\varrho\) and \(\vec v\) are no longer dynamical variables in the
Lagrangian picture; \(\vec f\) attains the status of the only
dynamical variable, if we demand that \(p\) is a functional of \(\vec
f\), which is obviously true for the assumption \(p=\alpha(\varrho)\).

Applying the first integral (\ref{varrho0varrhoJ}) and the
transformation of the Eulerian differentiation (\ref{transformation})
to the equations for \(\vec g\) (\ref{EulerNewtonc},
\ref{EulerNewtond}) we obtain the Lagrange--Newton--System with
pressure, i.e. a set of four coupled nonlinear partial differential
equations for the trajectory field \(\vec f\).  We first obtain:
\begin{subequations}
\begin{eqnarray}\nonumber
0 = (\vec\nabla \times \vec g )_{h} &=& (\vec\nabla \times \ddot{\vec f} )_{h}+
\frac{1}{\varrho ^{2}}(\vec\nabla p\times \vec\nabla \varrho )_{h}\\
\nonumber &=&  
\varepsilon _{hji}{\ddot f}_{i,j}+\frac{\varepsilon _{hji}}{\varrho ^{2}}
\varrho _{,i}p_{,j}\\
 &=&  \frac{1}{J}{\cal J} (\ddot f _{i},f_{i},f_{h})\nonumber
\\&+&\frac{1}{2J^{2}
\varrho ^{2}}\varepsilon _{jpq}{\cal J} (\varrho ,f_{h},f_{j}){\cal J} 
(p,f_{p},f_{q})\, ,\label{LagrangeNewton1a} \\ 
\Lambda -4\pi G\varrho& =&  \vec\nabla \cdot \vec g =\vec\nabla \cdot 
\ddot{\vec f} +\vec\nabla \cdot \frac{\vec\nabla p}{\varrho }\nonumber \\
&=& \frac{1}{2J}\varepsilon _{jpq}{\cal J} (\ddot f _{j},f_{p},f_{q})\nonumber
\\&+&
\frac{1}{2J}\varepsilon_{jpq}{\cal J}(\frac{1}{2J\varrho }\varepsilon _{jkl}
{\cal J} (p,f_{k},f_{l}),f_{p},f_{q})\nonumber \\
&=& \frac{1}{2J}\varepsilon _{jpq}{\cal J} (\ddot f _{j},f_{p},f_{q})\nonumber
\\&-&
\frac{1}{2J^{3}\varrho }{\cal J} (p,f_{p},f_{q}){\cal J} (J,f_{p},f_{q})
\nonumber \\
& -&\frac{1}{2J^{2}\varrho^{2}}{\cal J} (p,f_{p},f_{q}){\cal J} 
(\varrho ,f_{p},f_{q})\nonumber
\\&+&\frac{1}{2J^{2}\varrho }{\cal J} 
({\cal J} (p,f_{p},f_{q}),f_{p},f_{q})\,.  \label{LagrangeNewton1b}
\end{eqnarray}
\end{subequations}
In the special case \(p=\alpha (\varrho) \) we get our final set of
equations:
\begin{subequations}
\begin{eqnarray}
0 =  (\vec\nabla \times \ddot{\vec f} )_{h}&=&\varepsilon _{hji}
\ddot f _{i,j}=\frac{1}{J}{\cal J}(\ddot f _{i},f_{i},f_{h})\quad ,\label{LagrangeNewton2a}\\
\Lambda -4\pi G\varrho & =&  \frac{1}{2J}\varepsilon _{jpq}{\cal J}
(\ddot f _{j},f_{p},f_{q})\nonumber
\\&-&\frac{\alpha'}{2J^{3}\varrho}{\cal J}
(\varrho,f_{p},f_{q}){\cal J}(J,f_{p},f_{q})\nonumber \\
&-& \frac{\alpha'}{2J^{2}\varrho^{2}}{\cal J}(\varrho ,f_{p},f_{q}){\cal J}
(\varrho,f_{p},f_{q})\nonumber\\
&+& \frac{\alpha'}{2J^{2}\varrho}{\cal J}
({\cal J}(\varrho ,f_{p},f_{q}),f_{p},f_{q})\nonumber\\
&+&\frac{\alpha ''}{2J^{2}\varrho}{\cal J}(\varrho ,f_{p},f_{q}){\cal J}
(\varrho ,f_{p},f_{q})\quad  \label{LagrangeNewton2b}.
\end{eqnarray}
\end{subequations}
This system of equations can be closed with the help of the exact
integral (\ref{varrho0varrhoJ}). 

\subsection{Annotations}

Substituting the integral (\ref{varrho0varrhoJ}) into
(\ref{LagrangeNewton2b}) we obtain:
\begin{eqnarray}
\Lambda -4\pi G\frac {C}{J} & = & \frac {1}{2J}\varepsilon _{jpq}{\cal J}
(\ddot f _{j},f_{p},f_{q})\nonumber \\
 &+& \frac{\alpha '}{2CJ^{2}}{\cal J}({\cal J}(C,f_{p},f_{q}),f_{p},f_{q})
\nonumber \\
 &-& \frac{\alpha '}{2J^{3}}{\cal J}({\cal J}(J,f_{p},f_{q}),f_{p},f_{q})\nonumber \\
 &+& \frac{1}{2CJ^{3}}{\cal J}(C,f_{p},f_{q}){\cal J}(C,f_{p},f_{q})\left( \alpha ''-\frac{J}{C}\alpha '\right) \nonumber \\
 &-& \frac{1}{2J^{4}}{\cal J}(C,f_{p},f_{q}){\cal J}(J,f_{p},f_{q})\left( 2\alpha ''+\frac{J}{C}\alpha '\right) \nonumber \\
\label{DwithJandC}  
&+& \frac{C}{2J^{5}}{\cal J}(J,f_{p},f_{q}){\cal J}(J,f_{p},f_{q})\left( 
\alpha ''+\frac{2J}{C}\alpha '\right) \,.\nonumber \\&&
\end{eqnarray}
Let us now introduce the following approximation which is common in
cosmology: consider the definition \( \varrho (\vec X)=\varrho
_{\mathrm{H}}(1+\delta(\vec X) ) \), where \( \varrho _{\mathrm{H}} \)
denotes the background density of the mean matter distribution and \(
\delta \) the density contrast. At the time \( t=t_{0} \) we have \(
\varrho _{0}(\vec X)=\varrho_{\mathrm{H}0}(1+\delta _{0}(\vec X)) \).
As the density contrast is numerically very small at the time of
recombination (i.e. \( t_{0}=t_{\mathrm rec} \)), \( \varrho _{0}=\varrho
_{\mathrm{H}0} \) is a useful approximation.  Therefore we get \(
C_{|i}=(\varrho_{0}J_{0})_{|i}=:C_{{\mathrm H}|i}=0\) (for an
alternative exact argumentation see Appendix A), and we use
\(C_{\mathrm{H}}\) instead of \(C\) furtheron to emphasize its
homogeneity. The Lagrange--Newton--System then may be written in a
much simpler form, since all functional determinants with \( C \)
disappear:
\begin{subequations}
\begin{eqnarray}
{\cal J}(\ddot f _{i},f_{i},f_{h}) & = &  0\quad ,\label{LagrangeNewton3a}\\
\Lambda -4\pi G \frac{C_{\mathrm{H}}}{J} & = & \frac{1}{2J}\varepsilon _{jpq}
{\cal J} (\ddot f _{j},f_{p},f_{q})\nonumber
\\&-&\frac{\alpha '}{2J^{3}}{\cal J} ({\cal J} 
(J,f_{p},f_{q}),f_{p},f_{q})\nonumber \\
&+& \frac{C_{\mathrm{H}}}{2J^{5}}{\cal J} (J,f_{p},f_{q}){\cal J} 
(J,f_{p},f_{q})\left( \alpha ''+\frac{2J}{C_{\mathrm{H}}}\alpha '\right) \, .
\nonumber \\&& 
\label{LagrangeNewton3b}
\end{eqnarray}
\end{subequations}
Looking at Eq. (\ref{LagrangeNewton2b}) we conclude
that a (not necessarily physically relevant) possibility that
simplifies Eq. (\ref{LagrangeNewton3b}) further is the
following:
\begin{eqnarray}
\label{specialcase} \alpha '' & = & -\frac{2J}{C_{\mathrm{H}}}\alpha '
=-\frac{2}{\varrho }\alpha '\quad , \quad 
{\mathrm{i.e.}}\quad , \quad p = c_{1}\frac{1}{\varrho }+c_{2}\quad .
\end{eqnarray}
This ``equation of state'' plays a special role as will become clear
in Subsect.3.2.

\subsection{Reduction to planar symmetry}

In order to learn more about the structure of Eqs. (\ref{LagrangeNewton2a}, 
\ref{LagrangeNewton2b}) it is useful to look
at restrictions in symmetry: we restrict \( \vec x =\vec f (\vec X ,t)
\) to planar symmetry, i.e.,\footnote{In this subsection we use \(J_0=1\), 
hence \(C(\vec X)=\varrho_0(\vec X)\).}

\begin{equation}
\label{def1D} 
f_{1}=f_{1}(X_{1},t)\;\;\;\,\;\;\;  f_{2}=X_{2}\;\;\; ,\;\;\; 
 f_{3}=X_{3} \;\;\;\;; 
\end{equation}
the only direction in which motion takes place is the \( X_{1} \)--direction; 
\( f_{2} \) and \( f_{3} \) are
constant. Consequently, there are no velocities and accelerations in
the \( X_{2} \)-- and \( X_{3} \)--directions (in particular \(
g_{2}=g_{3}=0 \)). So, \( J \) simplifies to
\[ J=\frac{\partial x_{1}}{\partial X_{1}}=f_{1|1}\quad .\]
The four field equations that we have in 3D are reduced to the
equation for the divergence of \( g_1\),
\[
\Lambda -4\pi G\varrho =\Lambda -4\pi
G\frac{\varrho_0}{f_{1|1}}=g_{i,i}=\frac{g_{1|1}}{f_{1|1}}\quad ;\]
the curl of \(\vec g\) vanishes identically.  With the help of
\[
g_{1}=b_{1}+\frac{p_{,1}}{\varrho }=\ddot f _{1}+\frac{p_{|1}}{\varrho
  f_{1|1}}=\ddot f _{1}+\frac{p_{|1}}{\varrho _{0}}\] we get
\[
\Lambda f_{1|1}-4\pi G\varrho_0=\ddot f
_{1|1}+(\frac{p_{|1}}{\varrho_0})_{|1}\quad .\] Let us define \(
G_{1|1}:=\Lambda -4\pi G\varrho_0 \) (\( G(X_{1}):=g(X_{1},t_{0}) \))
in order to cast the last equation into a total divergence,
\[
( \ddot f _{1}-\Lambda f_{1}-G_{1}+\Lambda
X_{1}+\frac{p_{|1}}{\varrho_0})_{|1}=0\quad ;\] hence,
\[
\ddot f _{1}-\Lambda f_{1}=G_{1}-\Lambda
X_{1}-\frac{p_{|1}}{\varrho_0}\quad ,\] neglecting an irrelevant
function of time. With the relation \( p=\alpha (\varrho ) \) the
pressure--force term becomes
\[
\frac{p_{|1}}{\varrho_0}=\frac{\alpha '}{\varrho_0}\left(
  \frac{\varrho _{0}}{f_{1|1}}\right) _{|1}=\alpha '\left(
  \frac{G_{1|11}}{(G_{1|1}-\Lambda
    )f_{1|1}}-\frac{f_{1|11}}{f_{1|1}^{2}}\right) \quad .\] Altogether
we arrive at
\begin{equation}
\label{result1D1}
\ddot f _{1}-\Lambda f_{1}=G_{1}-\Lambda X_{1}+\alpha '\left( \frac{f_{1|11}}{f_{1|1}^{2}}-\frac{G_{1|11}}{(G_{1|1}-\Lambda )f_{1|1}}\right) \quad , 
\end{equation}
or (with ignorance of dimensions)
\begin{equation}
\label{result1D2}
\displaystyle \ddot f _{1}-\Lambda f_{1}=G_{1}-\Lambda X_{1}+\frac{\alpha '}{f_{1\vert 1}}\left( \ln \left( \frac{f_{1|1}}{G_{1\vert 1}-\Lambda }\right) \right) _{\vert 1}\quad . 
\end{equation}
For the case \(\Lambda =0 \) and \(\alpha '=const. \) G\"otz (1988)
has shown that (\ref{result1D2}) can be mapped to the
Sine--Gordon equation, which is a well--studied equation that admits
soliton solutions.

\section{Lagrangian perturbation approach}

\subsection{The perturbation ansatz and linearization}

In standard cosmology we invoke a homogeneous deformation of the
continuum: \(f_{{\mathrm{H}}i}=a_{ij}(t)X_{j}\) that is isotropic: \(
a_{ij}(t)=a(t)\delta _{ij} \); thus, with \( \vec f
_{\mathrm{H}}=a(t)\vec X \) and the homogeneous density \( \varrho
_{\mathrm{H}}:=\frac{C_{\mathrm{H}}}{a^{3}}\) applied to the
Lagrange--Newton--System with pressure we are left with
\begin{equation}
\label{fastFriedmann}
3\frac{\ddot a }{a}=\Lambda -4\pi G\frac{C_{\mathrm{H}}}{a^{3}}\quad .
\end{equation}
The first integral of this equation is Friedmann's equation:
\begin{equation}
\label{FriedmannscheDgl}
\frac{\dot a ^{2}+const.}{a^{2}}=\frac{8\pi G\frac{C_{\mathrm{H}}}{a^{3}}+\Lambda }{3}\quad .
\end{equation}
In order to describe structure formation we consider small (\( {\cal
  O}(\varepsilon) \)) deviations \( \vec p (\vec X ,t)\) from this
homogeneous and isotropic motion; we use the ansatz
\begin{eqnarray}
\label{perturbationansatz} \vec f (\vec X ,t) & = & a(t)\vec X +\vec p (\vec X ,t)\quad,
\end{eqnarray}
and suppose that perturbation theory is justifiable.  The assumption
that the perturbations are smooth leads us to \( |p_{i|j}|\ll a \) and
\( |\ddot p _{i|j}|\ll \ddot a \).  Inserting our ansatz
(\ref{perturbationansatz}) into the first equations
(\ref{LagrangeNewton3a}) we have
\begin{eqnarray}
0&=&(\vec\nabla \times \vec g )_{h}={\cal J} (\ddot f _{i},f_{i},f_{h}) 
\nonumber \\&=& {\cal J} (\ddot a X_{i}+\ddot p _{i},aX_{i}+p_{i},aX_{h}+p_{h})\nonumber \\
 & = &\ddot a a{\cal J} (X_{i},p_{i},X_{h})+\ddot a {\cal J} (X_{i},p_{i},p_{h})+a^{2}{\cal J} (\ddot p _{i},X_{i},X_{h})\nonumber\\
\label{perrot}&+& a{\cal J} (\ddot p _{i},X_{i},p_{h})+a{\cal J} 
(\ddot p _{i},p_{i},X_{h})+{\cal J} (\ddot p _{i},p_{i},p_{h}) \, .
\end{eqnarray}
In the first--order approximation we neglect quadratic and cubic
terms, hence:
\begin{subequations}
\begin{eqnarray}
\ddot a a{\cal J} (X_{i},p_{i},X_{h})+a^{2}{\cal J} (\ddot p _{i},X_{i},X_{h})  
&=&0  \;\;\mathrm{or}\;\; \mathrm{equivalently,} \nonumber \\
\vec\nabla _{\vec 0}\times \left(\ddot{\vec p} -\frac{\ddot a }{a}
\vec p\right)& =&  \vec 0 \quad .\label{perturbationa}
\end{eqnarray}
\end{subequations}

Linearizing the equation for the divergence of \(\vec g\) we start
with (\ref{LagrangeNewton3b}), i.e., with the assumption \( C =
C_{\mathrm{H}}\) (compare Appendix A).  The following terms appear
during the calculation:
\begin{eqnarray*}
J & = &  a^{3}+a^{2}p_{i|i}+{\cal O}(\varepsilon^{2}) \\
\frac{1}{J} & = &  \frac{1}{a^{3}(1+\frac{1}{a}p_{i|i})+
{\cal O}(\varepsilon^{2})}\\ &=&\frac{1}{a^{3}}(1-\frac{1}{a}p_{i|i})+
{\cal O}(\varepsilon^{2}) \\
\frac{1}{J^{n}} & = &  \frac{1}{a^{3n}}(1-\frac{n}{a}p_{i|i})+
{\cal O}(\varepsilon ^{2})\, \\
\varepsilon _{jpq}{\cal J} (\ddot f _{j},f_{p},f_{q}) & = &  6\ddot a a^{2}+4\ddot a ap_{i|i}+2a^{2}\ddot p _{i|i}+{\cal O}(\varepsilon^{2}) \\
{\cal J} (J,f_{p},f_{q}) & = &  \varepsilon _{apq}a^{4}p_{i|ia}+{\cal O}(\varepsilon^{2}) \\
{\cal J} ({\cal J} (J,f_{p},f_{q}),f_{p},f_{q}) & = &  2a^{6}p_{i|ijj}+{\cal O}(\varepsilon^{2}) \\
{\cal J} (J,f_{p}f_{q}){\cal J} (J,f_{p},f_{q}) & = &  {\cal O}(\varepsilon^{2}) 
\end{eqnarray*}
(remember: \( \varepsilon _{abc}\varepsilon _{abc}=6 \) and \(
\varepsilon _{abc}\varepsilon _{abd}=2\delta _{cd} \)). In the end we
find: \subnumbers
\begin{gather}
  \stepsubnumber 0 = -\Lambda+4\pi G\frac{ C_{\mathrm{H}}}{a^{3}}-4\pi
  G\frac{ C_{\mathrm{H}}}{a^{4}}p_{i|i}+3\frac{\ddot a}{a}-
  \frac{\ddot a}{a^{2}}p_{i|i}+\frac{1}{a}\ddot
  p_{i|i}
\nonumber \\
-\frac{\alpha'}{a^{3}}p_{i|ijj}+{\cal O}(\varepsilon^{2})
  \quad . \label{perturbationb}
\end{gather}
\nosubnumbers (The calculation without using \(C = C_{\mathrm H}\) may be found
in (Adler 1998).)  In comparison to the matter model ``dust'' (Buchert 1992) there is
only one additional term \( \frac{\alpha
  '}{a^{3}}p_{i|ijj}=\frac{\alpha '}{a^{3}}\Delta_{\vec 0}(p_{i|i}) \).  
Inserting the zero--order equation
(\ref{fastFriedmann}) (thus neglecting backreaction of the
first--order displacements on the background solution) yields
\[
0=-4\pi G\frac{ C_{\mathrm H}}{a^{4}}p_{i|i}-\frac{\ddot
  a}{a^{2}}p_{i\vert i}+\frac{1}{a}\ddot p_{i|i}-\frac{\alpha
  '}{a^{3}}p_{i|ijj}\quad .\]
Introducing the comoving displacement field \( \vec P:
=\frac{\vec p }{a(t)} \) representing the inhomogeneous deformation
that is scaled with the expansion, we get:
\[
0=-4\pi G\frac{ C_{\mathrm H}}{a^{3}}P_{i|i}+\ddot P_{i|i}+2\frac{\dot
  a }{a}\dot P_{i|i}-\frac{\alpha '}{a^{2}}P_{i|ijj}\quad .\] This
equation can be written as a total divergence:
\[
\vec\nabla _{\vec 0}\cdot \left( \ddot{\vec P} +2\frac{\dot a }{a}\dot{\vec
    P} - 4\pi G\frac{ C_{\mathrm{H}}}{a^{3}}\vec P -\frac{\alpha
    '}{a^{2}}\Delta _{0}\vec P \right) =0\quad .\] 
Writing Equation (\ref{perturbationa}) also in terms of the scaled
displacement field, and also inserting the zero--order solution, we arrive
at our main result:

\noindent
For vanishing harmonic parts (see Ehlers \& Buchert 1997) we obtain the
following final set of equations that comprises
the linear perturbation theory for the scaled displacement field 
\(\vec P=\vec P^L+\vec P^T\) \((\vec\nabla_{\vec 0}\times\vec P^L = \vec 0\;\);\(\;
\vec\nabla_{\vec 0}\cdot\vec P^T = 0\;)\;\):
\begin{subequations}
\begin{eqnarray}
\ddot{\vec P^T}+2\frac{\dot a}{a}\dot{\vec P^T}&=&\vec 0 \label{Pepsia}\\
\ddot{\vec P^L} +2\frac{\dot a}{a}\dot{\vec P^L} 
-4\pi G\frac{C_{\mathrm{H}}}{a^{3}}
\vec P^L &=&\frac{\alpha '}{a^{2}}\vec\Delta _{\vec 0}\vec P^L \quad . \label{Pepsib}
\end{eqnarray}
\end{subequations}
Remark:

\noindent
Remember that \(\alpha'=\alpha'(\varrho)\) where \(\varrho\) depends
on the deformation \(\vec P\); thus, the r.h.s. of Eq. (\ref{Pepsib})
is not linear in \(\vec P^L\). So, for given \(p=\alpha (\varrho)\),
\(\alpha' \vec\Delta_{\vec 0}\vec P^L\) has to be linearized with respect to
\(\vec P^L\) in the context of linear perturbation theory, i.e.
\(\alpha' = const\).  This fact amounts to a major difference compared
to the ``dust'' case: in the latter, the dependence on \(\varrho \)
completely disappears in the equations and the exact density may be
obtained for any perturbative solution. Here, the density is
implicitly linearized and this fact has already been used during the
above derivation.

\subsection{Comparison with plane--symmetric inhomogeneities\\
  on a Friedmann background universe}

Let us start with trajectories, which are scaled with the expansion
\(\vec F=\frac {\vec f}{a(t)}\).  Reduction to planar symmetry now
means:
\begin{equation}
F_{1} = X_{1}+P_{1}(X_{1},t)\label{Definition3.2}\,\, ,\,
F_{2} = X_{2}\,\, , \,
F_{3} = X_{3}\quad .
\end{equation}
The functional determinant \( J \) reduces to
\[
J=a^{3}(1+P_{1|1})\quad ,\] and, thus, \( J \) is independent of the
coordinates \( X_{2} \) und \( X_{3} \).
With the simplification \( C =C_{\mathrm{H}} \) we derive from
(\ref{LagrangeNewton3b})
\begin{eqnarray*}
 \Lambda -4\pi G\frac{C_{\mathrm{H}}}{J}& = & \frac{ 2a^{2}\ddot a}{J} 
(1+P _{1|1})\\&+&\frac{a^{2}}{J}(\ddot a (1+P_{1|1})+2\dot a \dot P _{1|1}+
a\ddot P _{1|1})\\
 &-&  \frac{\alpha '}{J^{3}}a^{4}J_{|11}+\frac{C _{\mathrm{H}}}{J^{5}}
(\alpha ''+\frac{2J}{C_{\mathrm{H}}}\alpha ')a^{4}J_{|1}J_{|1}\, ,
\end{eqnarray*}
i.e.,
\begin{eqnarray*}
\Lambda &-& 4\pi G\frac{C_{\mathrm{H}}}{a^{3}(1+P_{1|1})} =  \\
&&3\frac{\ddot a }{a}+2\frac{\dot a }{a}\frac{\dot P _{1|1}}{(1+P_{1|1})}
+\frac{\ddot P _{1|1}}{1+P_{1|1}} 
- \frac{\alpha '}{a^{2}(1+P_{1|1})^{3}}P_{1|111} \\
&&+ \frac{C_{\mathrm{H}}}{a^{5}(1+P_{1|1})^{5}}
(\alpha'' +\frac{2(1+P_{1|1})a^{3}}{C_{\mathrm{H}}}\alpha ')
P_{1|11}P_{1|11}\, .
\end{eqnarray*}
Using the zero--order equation (\ref{fastFriedmann}) and multiplying
with \((1+P_{1|1})\) yields
\begin{eqnarray*}
\left(\ddot{P _{1}}+2\frac{\dot a }{a}\dot P _{1} -4\pi G \frac{C_{\mathrm{H}}}{a^{3}}P _{1} \right) _{|1} =  \frac{\alpha '}{a^{2}(1+P_{1|1})^{2}}P_{1|111}\\
 - \frac{C_{\mathrm{H}}}{a^{5}(1+P_{1|1})^{4}}(\alpha ''+\frac{2(1+P_{1|1})a^{3}}{C_{\mathrm{H}}}\alpha ')P_{1|11}P_{1|11}\quad . 
\end{eqnarray*}
The r.h.s. of this equation can be written as a total derivative;
hence we get for vanishing harmonic parts,
\begin{equation}
\label{4.3end} \displaystyle \ddot{P _{1}}+2\frac{\dot a }{a}\dot P _{1} -
4\pi G \frac{C_{\mathrm{H}}}{a^{3}}P _{1}= \frac{\alpha '}{a^{2}(1+P_{1|1})^{2}}P_{1|11} \quad .
\end{equation}
By linearizing this last equation we recover Eq. (\ref{Pepsib})
for one-dimensional inhomogeneities.  Writing the coefficient on the
r.h.s. as a function of \(\varrho \),
\begin{displaymath}
\chi (\varrho ) =\frac{\alpha '(\varrho )}{a^{2}(1+P_{1|1})^{2}}= 
\frac{\alpha'(\varrho)}{a^2}\left(\frac{\varrho}{\varrho_{\mathrm{H}}}\right)^2 \quad ,
\end{displaymath} 
we see that we get a linear equation in the case of the special
``equation of state'' (\ref{specialcase}).

Let us finally restrict our problem to the limit of no expansion
\(a(t)=a(t_{0})\). We have \(\varrho_{\mathrm H}= C_{\mathrm H}\) 
and \(f_{1}=a(t_{0})(X_{1}+P_{1})\) from
(\ref{Definition3.2}). Therefore,
\begin{displaymath}
\ddot{f _{1}}-4\pi G\frac{C_{\mathrm{H}}}{a(t_{0})^3}
(f_{1}-a(t_0)X_{1})=\frac{\alpha '}{f_{1|1}^{2}}f_{1|11}\quad .
\end{displaymath}
If we additionally appreciate that (\ref{fastFriedmann}) now reads
\(\Lambda =4\pi G C_{\mathrm H}\) and \(G_{1|1}=\Lambda -4\pi G 
\varrho_{{\mathrm H}0} = 0\), we recover Eq. (\ref{result1D2})  
in the limit of the static Einstein universe.

\subsection{Comparison with Eulerian linear theory}

It is a standard procedure in cosmology to scale the Eulerian coordinates with the
homogeneous and isotropic expansion (\( \vec v _{\mathrm{H}}=H\vec x
\) with Hubble's function \(H:=\frac {\dot a}{a} \)), i.e. to use the
following transformation of coordinates
\begin{equation}
\label{qQ} 
\vec q =\vec Q (\vec x ,t)=\frac{\vec x }{a(t)} \quad .
\end{equation}
The comoving differential operators \( (\vec\nabla _{\vec q
  })_{i}=\frac{\partial}{\partial q_{i}} \)\, and \( \frac{\partial
  }{\partial t}\mid _{\vec q } \) become
\begin{eqnarray*}
&&\vec\nabla  = \frac{1}{a(t)}\vec\nabla _{\vec q }\\
&&\frac{\partial }{\partial t}\Big \vert _{\vec x } =  
\frac{\partial }{\partial t}\Big \vert _{\vec q }-H\vec q\, \vec\nabla _{\vec q }\quad .
\end{eqnarray*}
The deviations from the homogeneous fields are called
\begin{eqnarray}
\vec u (\vec q ,t) & := & \vec v (\vec q ,t)-\vec v _{\mathrm{H}}(\vec q ,t)\quad \: \mathrm{peculiar}\, \mathrm{velocity},\nonumber \\
\vec w (\vec q ,t) & := & \vec g (\vec q ,t)-\vec g _{\mathrm{H}}(\vec q ,t)\quad \: \mathrm{peculiar}\, \mathrm{acceleration}\: \mathrm{and}\nonumber \\
\delta (\vec q ,t) & := & \frac{\varrho (\vec q ,t)-\varrho _{\mathrm{H}}(t)}{\varrho _{\mathrm{H}}(t)}\quad \quad \mathrm{density}\, \mathrm{contrast},\nonumber 
\end{eqnarray}
where \( {\vec v}_{\mathrm H} = \dot a \vec q \)\, and \( {\vec g}_{\mathrm H}=
\ddot a \vec q \) denote the Hubble--velocity and the Hubble--acceleration. We
obtain the Euler--Newton--System with pressure for the peculiar
quantities by making use of the homogeneous solutions (see Peebles
1980):
\begin{subequations}
\begin{eqnarray}
\dot\delta +\frac{1}{a}(1+\delta )\vec\nabla _{\vec q }\cdot \vec u  & =&  0\quad ,\label{noneeda}\\
\dot{\vec u} +H\vec u  & =&  \vec w -\frac{1}{a}\frac{\vec\nabla _{\vec q }\, p}{\varrho _{\mathrm{H}}(1+\delta )}\quad ,\label{noneedb}\\
\vec\nabla _{\vec q }\times \vec w &  =&  \vec 0 \quad ,\label{noneedc}\\
\vec\nabla _{\vec q }\cdot \vec w   &= & -4\pi G\varrho _{\mathrm{H}}\delta a\quad .\label{noneedd}
\end{eqnarray}
\end{subequations}
In the following we assume that all peculiar quantities and their
derivatives are small (\( {\cal O}(\varepsilon) \)). Using
perturbation theory to the first order in \(\varepsilon\) we get the
linearized form of the equations:
\begin{subequations}
\begin{eqnarray}
\frac{\partial }{\partial t}\Big \vert _{\vec q }\delta +\frac{1}{a}\vec\nabla _{\vec q }\cdot \vec u  & =&  {\cal O}(\varepsilon^{2})\label{eulerlina} \\
\frac{\partial }{\partial t}\Big \vert _{\vec q }\vec u +H\vec u  & =&  \vec w -\frac{1}{a}\frac{\vec\nabla _{\vec q }\, p}{\varrho _{\mathrm{H}}}+{\cal O}(\varepsilon^{2}) \label{eulerlinb}\\
\vec\nabla _{\vec q }\times \vec w &  = & \vec 0 \label{eulerlinc}\\
\vec\nabla _{\vec q }\cdot \vec w  & = 
& -4\pi G\varrho _{\mathrm{H}}\delta a\quad .
\label{eulerlind}
\end{eqnarray}
\end{subequations}
The linearized equation for the evolution of \( \delta \) can be
calculated by applying \( -\frac{1}{a}\vec\nabla _{\vec q } \) to the
second equation and inserting the first and forth equations:
\begin{equation}
\label{delta}  \displaystyle \frac{\partial ^{2}}{\partial t^{2}}\Big 
\vert _{\vec q }\delta +2{\mathrm{H}}\frac{\partial }{\partial t}\Big 
\vert _{\vec q }\delta -4\pi G\varrho _{\mathrm{H}}\delta =
\frac{\Delta _{\vec q }\, p}{a^{2}\varrho _{\mathrm{H}}}=
\left(\frac{c_{s}}{a}\right)^{2}\Delta _{\vec q }\, \delta \quad ;
\end{equation}
where we have set \( p=\alpha (\varrho)\), and \( \alpha '=\frac{\partial p}{\partial 
\varrho }=:c_{s}^{2} \)
is defined in terms of the ``speed of sound'' \(c_s\) as is usual in a
hydrodynamical medium. Here, we have \(\alpha' = const.\), i.e., \( p \propto
\varrho \), in order to obtain a linear equation in \(\delta\).  
\bigskip

\section{Discussion of Results and Outlook}

\addtocounter{equation}{1} In the last section we have derived three
different equations:
\begin{itemize}
\item From the linearization of the Lagrange--Newton--System with
  pressure we have for the longitudinal part of the inhomogeneous
  deformation \( \vec P (\vec X ,t)=\frac{\vec p (\vec X ,t)}{a(t)}
  \):
\begin{gather}
  \ddot{\vec P^L} +2\frac{\dot a }{a}\dot{\vec P^L} -4\pi G\varrho
  _{\mathrm{H}}\vec P^L =\frac{\alpha '}{a^{2}}\vec\Delta_{\vec 0}\vec P^L
  \quad ;\label{A} \tag{\theequation $\mathrm{A}$}
\end{gather}
remember \(\varrho_{\mathrm{H}}:=\frac{C_{\mathrm{H}}}{a^{3}}\).
\item The plane--symmetric solutions on a Friedmannian background
  obey:
\begin{gather}
  \ddot{P _{1}}+2\frac{\dot a }{a}\dot P _{1} -4\pi G \varrho
  _{\mathrm{H}}P _{1}= \frac{\alpha '}{a^{2}(1+P_{1|1})^{2}}P_{1|11}
  \quad .\label{B} \tag{\theequation $\mathrm{B}$}
\end{gather}
\item The Eulerian linear theory in comoving coordinates \(\vec
  q=\frac{\vec x}{a(t)}\) leads to a partial differential equation for the
  density contrast \( \delta (\vec q ,t) \):
\begin{gather}
  \frac{\partial ^{2}}{\partial t^{2}}\Big \vert _{\vec q }\delta
  +2\frac{\dot a }{a}\frac{\partial }{\partial t}\Big \vert _{\vec q
    }\delta -4\pi G\varrho _{\mathrm{H}}\delta
  =\frac{\alpha'}{a^{2}}\Delta _{\vec q }\delta \quad .\label{C}
  \tag{\theequation $\mathrm{C}$}
\end{gather}
\end{itemize}
According to the equivalence of the equations (\ref{A}) and (\ref{C})
up to the time derivative operators \(
\frac{\mathrm{d}}{{\mathrm{d}}t}=\frac{\partial }{\partial t}\vert
_{\vec X } \) and \ \( \frac{\partial }{\partial t}\vert _{\vec q }
\), we see that, with the help of the already known results for the
density contrast in the Eulerian linear theory (e.g. Haubold et al.
1991), solutions for the inhomogeneous deformation \( \vec P^L \) in
the Lagrangian linear theory can be constructed.  In the case \( p=0
\) a class of exact 3D solutions has been found with the help of this
method (see: Buchert 1989). But, in contrast to the case \( p=0 \),
this is not to be expected here, since extrapolation of the solution
of the linearized equations does not yield an exact solution of the
planar problem as well; this is easy to see from Eq.
(\ref{B}): the pressure term produces (except in the special case
(\ref{specialcase})) nonlinear terms in \( \vec P \) already for
plane--symmetric inhomogeneities.

We emphasize the special role played by the ``equation of state''
(\ref{specialcase}): \(p=\frac {c_1}{\varrho} +c_2\).

\smallskip

For further applications we can proceed as follows:

\noindent
We look at solutions \( \delta ^{\ell}(\vec q ,t) \) of the
differential equation (\ref{C}) and use the instruction \( \vec q \:
\longmapsto \: \vec X ,\: \delta ^{\ell}\longrightarrow
P_{i}^{\mathrm{L}} \) to construct solutions \(
P_{i}^{\mathrm{L}}(\vec X ,t) \) of the differential equations
(\ref{A}). As the Lagrangian description implicitly respects
nonlinearities, our construction rule allows to build nonlinear models
of structure formation that take dynamical pressure forces into
account.  Since solutions to Equation (\ref{A}) constitute the
Lagrangian extrapolation of the Eulerian linear theory, they are built
in the same spirit as Zel'dovich's approximation in the case \(p=0\)
(see: Zel'dovich 1970, 1973; Buchert 1989).  Viewed together with the
derivation of the ``adhesion approximation'' by Buchert \& Dom\'\i
nguez (1998), these solutions may be used as first approximations to
adhesive gravitational clustering in the weakly nonlinear regime.


\begin{acknowledgements}

TB is supported by the ``Sonderforschungsbereich SFB 375 f\"ur
Astro--Teilchenphysik der Deutschen Forschungsgemeinschaft''.  The
authors wish to thank Claus Beisbart and Alvaro Dom\'\i nguez for
valuable discussions and comments.

\end{acknowledgements}

\newpage

\section*{Appendix A: Alternative argument for the approximation \(C_{|i}=0\)}

We want to give an alternative argumentation for the approximation \(
C _{|i}=(\varrho_{0}J_{0})_{|i}=0 \).  Let us introduce new
curvilinear coordinates \(\vec Y\) by
\begin{equation}
\label{xytransformation}
\vec Y :=\vec A (\vec X )=\vec X +\vec\Psi (\vec X )\quad .
\end{equation}
Below we shall consider (initially) small deviations \(\vec \Psi \)
from \(\vec X\), restricting our argument to the linear approximation.
We choose the new coordinates in such a way that
\[C(\vec Y)=\varrho_{0}(\vec Y)J_0(\vec Y)=
const.=: C_{\mathrm{H}}\quad;\quad C_{\mathrm H}=\varrho_{\mathrm{H}0}\quad.\] 
Thus, taking the conservation of mass into account, we have
\[
{\mathrm{d}}m_0=\varrho _{0}(\vec X){\mathrm{d}^{3}}X=\varrho
{\mathrm{d}^{3}}Y \quad. \] Writing
\(\varrho_0(\vec X)=\varrho _{\mathrm{H}0}(1+\delta_0(\vec X))\), the
map \(\vec A\) is defined by
\[
(1+\delta_0(\vec X))\det \! \left( \frac{\partial X_{i}}{\partial
    Y_{j}}\right) =1\quad .\] Then we get for \( |\vec\Psi
|={\cal O({\varepsilon})} \) and small derivatives:
\[
1+\delta _{0}(\vec X )=\det \! \left( \frac{\partial
    Y_{j}}{\partial X_{i}}\right) =\det \! \left( \delta
  _{ij}+\frac{\partial \Psi _{j}}{\partial X_{i}}\right) \]
\[
= 1+\vec\nabla_{\vec 0}\cdot \vec\Psi (\vec X )+{\cal O}(\varepsilon^{2}) \quad .\] 
Thus,
to first order in \(\varepsilon \) we have:
\[
\delta_{0}(\vec X )=\vec\nabla _{\vec 0}\cdot \vec\Psi (\vec X )\quad ,\]
the particles are initially displaced according to the density
contrast.
With \( \vec\nabla _{\vec 0}\cdot \vec W =-4\pi G\varrho
_{\mathrm{H}0}\delta _{0}a(t_0) \) ((\ref{eulerlind}) at the time \(
t=t_{0}\)) \(\vec \Psi \) is defined with the help of the 
peculiar--acceleration \( \vec w \) at time \(t_0\) :
\begin{equation}
\label{BeziehungPsiW}
\vec\nabla _{\vec 0}\cdot \vec\Psi =-\frac{1}{4\pi G\varrho _{\mathrm{H}0}a(t_0)}
\vec\nabla _{\vec 0}\cdot \vec W \: ,\: \quad \vec W :=\vec w (\vec X ,t_{0})
\quad \;\;.
\end{equation}
To enforce the property \(C_{|i}=0\), we simply have to
relabel the trajectories:
\[ {\vec X} \longrightarrow \vec Y, \quad  \vec x=\vec f(\vec Y,t)\,\,{ \mathrm{with}}\,\,\,\vec x|_{t_0}=\vec Y=\vec X+\vec{\Psi}(\vec X)\quad.\]
Note that Zel'dovich's original discussion of his approximation
(Zel'dovich 1970, 1973) as well as subsequent work were all using the
coordinates \(\vec Y\). We stress, however, that the above
argument also involves an approximation and is only consistent within
the first--order solutions; compare the discussion in (Buchert 1989).

\section*{Appendix B: Transformed vector--identities}

The following equations comprise a useful collection of formulas;
similar expressions may arise in calculations employing Lagrangian
coordinates.  We list these transformed vector--identities here,
because they may be helpful for further considerations.
\begin{eqnarray}
0 & = &  (\vec\nabla \times (\vec\nabla k))_{h}=
\frac{1}{J}{\cal J} (k_{,i},f_{i},f_{h})\nonumber \\
\label{Vektorid1}  & = & \frac{1}{2J^{2}}\varepsilon _{ipq}{\cal J} 
({\cal J} (k,f_{p,}f_{q}),f_{i},f_{h})\nonumber \\
&-&\frac{1}{2J^{3}}\varepsilon _{ipq}{\cal J} 
(k,f_{p},f_{q}){\cal J} (J,f_{i},f_{h})\quad ;\nonumber \\
 &  & \nonumber \\
0 & = &  \vec\nabla \cdot (\vec\nabla \times \vec T )=
(\frac{1}{J}{\cal J} (T_{p},f_{p},f_{q}))_{,q}\nonumber \\
\label{Vektorid2}  & = & \frac{1}{2J^{2}}\varepsilon _{ihq}{\cal J} 
({\cal J} (T_{p},f_{p},f_{q}),f_{i},f_{h}) \nonumber \\
&-&\frac{1}{2J^{3}}\varepsilon _{ihq}{\cal J} 
(T_{p},f_{p},f_{q}){\cal J} (J,f_{i},f_{h})\quad ;\nonumber \\
 &  & \nonumber \\
0 & = & \left( \vec\nabla \times (\vec\nabla \times \vec T )-
\vec\nabla (\vec\nabla \cdot \vec T )+\Delta \vec T \right) _{h}\nonumber \\
 & = & (\frac{1}{J}\varepsilon_{hji}{\cal J} (T_{r},f_{r},f_{i}))_{,j}
\nonumber \\
&-&(\frac{1}{2J}\varepsilon_{jpq}{\cal J} (T_{j},f_{p},f_{q}))_{,h}
\nonumber \\
&+&(\frac{1}{2J}\varepsilon _{ipq}{\cal J} (T_{h},f_{p},f_{q}))_{,i}\nonumber 
\\
 & = & \frac{1}{J^{2}}{\cal J} ({\cal J} (T_{r},f_{r},f_{i}),f_{i},f_{h})
\nonumber \\
&-&\frac{1}{J^{3}}{\cal J} (T_{r},f_{r},f_{i}){\cal J} (J,f_{i},f_{h})\nonumber \\
&-& \frac{1}{4J^{2}}\varepsilon_{hrt}\varepsilon_{jpq}{\cal J} ({\cal J} 
(T_{j},f_{p},f_{q}),f_{r},f_{t})
\nonumber \\
&+&\frac{1}{4J^{3}}\varepsilon_{hrt}\varepsilon_{jpq}{\cal J} 
(J,f_{r},f_{t}){\cal J} (T_{j},f_{p},f_{q})\nonumber \\
\label{Vektorid3}  &+& \frac{1}{2J^{2}}{\cal J}({\cal J} 
(T_{h},f_{r},f_{t}),f_{r},f_{t})
\nonumber
\\&-&\frac{1}{2J^{3}}{\cal J} (J,f_{r},f_{t}){\cal J} (T_{h},f_{r},f_{t})\quad .\nonumber 
\end{eqnarray}

\end{document}